\documentclass[aps,amsfonts,amsmath,prd,preprint,nofootinbib]{revtex4-1}
\pdfoutput=1
\usepackage{graphicx}

\newcommand{\beq}{\begin{equation}}
\newcommand{\eeq}{\end{equation}}

\def\lap{\lower.5ex\hbox{$\; \buildrel < \over \sim \;$}}
\def\gap{\lower.5ex\hbox{$\; \buildrel > \over \sim \;$}}

\begin{document}

\title{Collapse of simple harmonic universe}

\author{Audrey T. Mithani and Alexander Vilenkin}

\address{
Institute of Cosmology, Department of Physics and Astronomy,\\ 
Tufts University, Medford, MA 02155, USA}

\begin{abstract}

In a recent paper Graham et al constructed oscillating and static universe models which are stable with respect to all classical perturbations.  Here we show that such universes are quantum-mechanically unstable and can collapse by quantum tunneling to zero radius.  We also present  instantons describing nucleation of oscillating and static universes from nothing.


\end{abstract}

\maketitle

\section{Introduction}

It has been recently shown \cite{Borde} that the spacetime of an inflationary universe is necessarily past-incomplete, even though inflation may be eternal to the future.  All past-directed timelike and null geodesics, except maybe a set of measure zero, reach the boundary of the inflating region of spacetime in a finite proper time (finite affine length, in the null case).  This indicates that inflation must have had some sort of a beginning.  One possibility is that the universe could have spontaneously nucleated out of nothing \cite{AV82}.

Unlike earlier singularity theorems, the theorem of Ref.~\cite{Borde} does not rely on Einstein's equations and does not assume any energy conditions.  To show the incompleteness of a given geodesic, all it requires is that the expansion rate averaged along the geodesic is greater than zero,\footnote{The expansion rate $H$ is defined in terms of a comoving congruence; see \cite{Borde} for details.}
\beq
H_{av} > 0.
\label{H>0}
\eeq
This is a rather weak condition, but it points to a possible loophole in the argument: the universe could be static in the asymptotic past.  Then $H_{av}=0$ and the theorem does not apply.  For example, a closed Friedmann-Robertson-Walker spacetime 
\beq
ds^2 = dt^2 - a^2(t) d\Omega_3^2
\label{metric}
\eeq
with a scale factor 
\beq
a(t) = a_0(1 + e^{H_0 t})
\eeq
and $a_0,H_0 = {\rm const}$ describes an inflating universe at $t>0$
and is geodesically complete both to the future and to the past.   The
idea that the universe could have started as a static, closed space in
the asymptotic past has been widely discussed in recent years, under
the name of ``emergent universe'' scenario (see, e.g.,
\cite{Mulryne,Sergio,Yu} and references therein). 

Construction of emergent universe models is a challenging task.  First
of all, it is not easy to arrange for a static universe to be stable.
It is well known that Einstein's static model, describing a closed
universe filled with nonrelativistic matter and positive vacuum
energy, is unstable with respect to small perturbations of the radius.
Even if the universe is initially perfectly fine-tuned, it will be
destabilized by quantum fluctuations and will either start inflating
or collapse to a singularity.  Such a universe cannot survive for an
infinite time.  However, 
radial stability can be achieved in models with modified gravity
\cite{Mulryne,Sergio,Yu} or with ``exotic'' matter
\cite{Dabrowski,Barrow,Graham} (see also \cite{Coule,Grisha}).   

A particularly simple model of the latter kind was recently discussed by Graham et al \cite{Graham}.  It describes a closed universe with a negative cosmological constant and a matter source with equation of state 
\beq
P=w\rho.
\label{P}
\eeq
The energy density of the universe is then
\beq
\rho = \Lambda + \rho_0 a^{-3(1+w)}
\label{rho}
\eeq
with $\Lambda<0$ and $\rho_0 >0$, and the Friedmann evolution equation is
\beq
{\dot a}^2 + 1 = \frac{8\pi G}{3} \rho a^2 .
\label{Friedmann}
\eeq
The model is radially stable, provided that $w$ satisfies
\beq
-1<w<-1/3.
\label{w}
\eeq
For a perfect fluid source, the speed of sound $c_s$ can be found from $c_s^2 = dP/d\rho = w$.  
With the equation of state (\ref{w}), this gives $c_s^2 < 0$,
indicating instability with respect to short-wavelength compressional
perturbations.  Hence, it is important that the exotic matter source
should not be a perfect fluid \cite{Graham}.\footnote{A perfect fluid
  source could be acceptable if one allows an equation of state
  $P(\rho)$ more general than Eq.~(\ref{P}).  All one needs is that
  $w=P/\rho$ satisfies Eq.~(\ref{w}) and $dP/d\rho >0$.}
  It could, for example,
be an assembly of randomly oriented domain walls, in which case
\cite{Spergel} $w=-2/3$ and $c_s^2 > 0$.  For this choice of $w$,
Eq.~(\ref{rho}) takes the form 
\beq
\rho = \Lambda + \rho_0 a^{-1},
\label{rhosimple}
\eeq
and the evolution equation (\ref{Friedmann}) has a simple oscillatory solution
\beq
a = \omega^{-1} (\gamma-\sqrt{\gamma^2 -1} \cos(\omega t))
\label{a}
\eeq
where 
\beq
\omega = \sqrt{\frac{8\pi}{3} G|\Lambda|}
\label{omega}
\eeq
and
\beq
\gamma=\sqrt{\frac{2\pi G\rho_0^2}{3|\Lambda|}}.
\label{gamma}
\eeq
Graham et al in Ref.~\cite{Graham} focused primarily on this special case, which they referred to as ``simple harmonic universe''.  

A static universe solution is obtained from (\ref{a}) by setting $\gamma =1 $; then $a = 1/\omega$.  It has been shown in \cite{Graham} that this solution is stable with respect to arbitrarily small perturbations, including all scalar and tensor modes.

Apart from stability, emergent universe models need a mechanism that would trigger inflationary expansion after an infinitely long stationary phase.  A possible mechanism has been suggested in \cite{Mulryne}.  It involves a massless scalar field $\phi$ with a self-interaction potential $V(\phi)$, such that $V\to {\rm const}$ at $\phi\to -\infty$.  In the stationary regime, the field ``rolls'' from $-\infty$ at a constant speed, ${\dot\phi}={\rm const}$.  Inflation is triggered when $\phi$ arrives at the non-flat region of the potential.  A similar mechanism has been employed in Ref.~\cite{Sergio}.

At the classical level, the simple harmonic universe model, supplemented with a suitable mechanism to trigger inflation, yields a consistent emergent universe scenario.  Our goal in this paper is to investigate whether or not the model remains stable in the quantum theory.  We shall restrict the analysis to the simplest minisuperspace model with a single dynamical degree of freedom -- the radius of the universe $a$.  We shall see that already at this simplest level the emergent universe exhibits a quantum instability.

\section{Collapse through tunneling}

We consider a spherical universe (\ref{metric}) with a matter content described by Eq.~(\ref{rho}).  The scale factor $a(t)$ is the single dynamical degree of freedom.  In classical theory, such a universe can be regarded as a constrained dynamical system with a Hamiltonian
\beq
{\cal H} = -\frac{G}{3\pi a}\left( p_a^2 + U(a) \right),
\label{H}
\eeq
where 
\beq
p_a = -\frac{3\pi}{2G}a{\dot a}
\eeq
is the momentum conjugate to $a$ and
\beq
U(a) = \left(\frac{3\pi}{2G}\right)^2 a^2\left(1-\frac{8\pi G}{3}a^2\rho(a)\right).
\label{U1}
\eeq
The Hamiltonian constraint ${\cal H}=0$ then yields the evolution equation (\ref{Friedmann}).

In quantum theory, the universe is described by a wave function $\psi(a)$, the conjugate momentum $p_a$ becomes the differential operator $-id/d a$ and the constraint is replaced by the Wheeler-DeWitt (WDW) equation \cite{DeWitt} (for a review see, e.g., \cite{AV94,Kiefer,Halliwell})
\beq
{\cal H}\psi=0,
\label{WDWeq}
\eeq 
or
\beq
\left(-\frac{d^2}{da^2} -\frac{\beta}{a}\frac{d}{da} +U(a)\right)\psi(a)=0.
\label{WDW}
\eeq
Here, the parameter $\beta$ represents the ambiguity in the ordering of the non-commuting factors $a$ and $p_a$ in the Hamiltonian (\ref{H}).  Its value does not affect the wave function in the semiclassical regime $a\gg l_{Planck}$.  From here on we set $\beta=0$.

One might expect that for a simple harmonic universe the potential $U(a)$ should be of the same form as for a harmonic oscillator.  This, however, is not the case: the motion in the potential (\ref{U1}) is simple harmonic only for a particular value of the energy, ${\cal H}=0$.  With $\rho(a)$ from (\ref{rhosimple}), we have
\beq
U(a) = \left(\frac{3\pi}{2G}\right)^2 a^2\left(1-\frac{8\pi G}{3}(\rho_0 a +\Lambda a^2) \right).
\label{U}
\eeq

It will be convenient to introduce a rescaled variable $x=\omega a$ with $\omega$ from Eq.~(\ref{omega}).  In terms of this variable the WDW equation takes the form
\beq
\left(-\frac{d^2}{dx^2} +U(x)\right)\psi(x)=0,
\label{WDWx}
\eeq
where
\beq
U(x) = \lambda^{-2} x^2 (1 - 2\gamma x + x^2),
\label{Ux}
\eeq
$\gamma$ is given by Eq.~(\ref{gamma}), and
\beq
\lambda = \frac{16 G^2|\Lambda|}{9}.
\eeq

The classically allowed range is defined by $U(x)\leq 0$.  This range is non-empty when $\gamma > 1$.  The shape of the potential in this case is illustrated in Fig.~1.  In the classical solution, the radius of the universe oscillates forever between the values $x_+$ and $x_-$ where $U(x_\pm) =0$,
\beq  
x_\pm = \gamma \pm \sqrt{\gamma^2 -1}.
\eeq
However, it is clear from the figure that quantum-mechanically the universe can tunnel through the barrier to a vanishing size at $x=0$.  The WKB tunneling action is given by
\beq
S = \int_0^{x_-} \sqrt{U(x)}dx  
\label{WKB}
\eeq
and the corresponding tunneling probability can be estimated as
\beq
{\cal P}\sim e^{-2S}.
\label{prob}
\eeq
This can be interpreted as the probability of collapse through quantum tunneling as the universe bounces at radius $x=x_-$.

Semiclassical quantum tunneling in oscillating universe models has been studied by Dabrowski and Larsen \cite{DabrowskiLarsen}.  They considered a closed universe containing nonrelativistic matter (dust), a domain wall fluid with equation of state $w=-2/3$, and a negative cosmological constant. 
Due to the presence of dust, this model has another classically allowed range at small values of $a$.  
The WKB action (\ref{WKB}) for tunneling between the two classically allowed regimes can then be expressed in terms of elliptic integrals.  In the absence of dust, the model of \cite{DabrowskiLarsen} reduces to the simple harmonic universe, but the authors have not discussed this case.

For a simple harmonic universe, the integral in (\ref{WKB}) can be expressed in terms of elementary functions,
\beq
S = \lambda^{-1} \left[ \frac{\gamma^2}{2} + \frac{\gamma}{4}\left( \gamma^2 -1 \right) \ln\left( \frac{\gamma-1}{\gamma+1} \right) -\frac{1}{3} \right],
\label{Sgeneral}
\eeq
Since the tunneling probability (\ref{prob}) is nonzero, such a universe cannot survive forever.
    
\begin{figure}[t]
\begin{center}
\includegraphics[width=12cm]{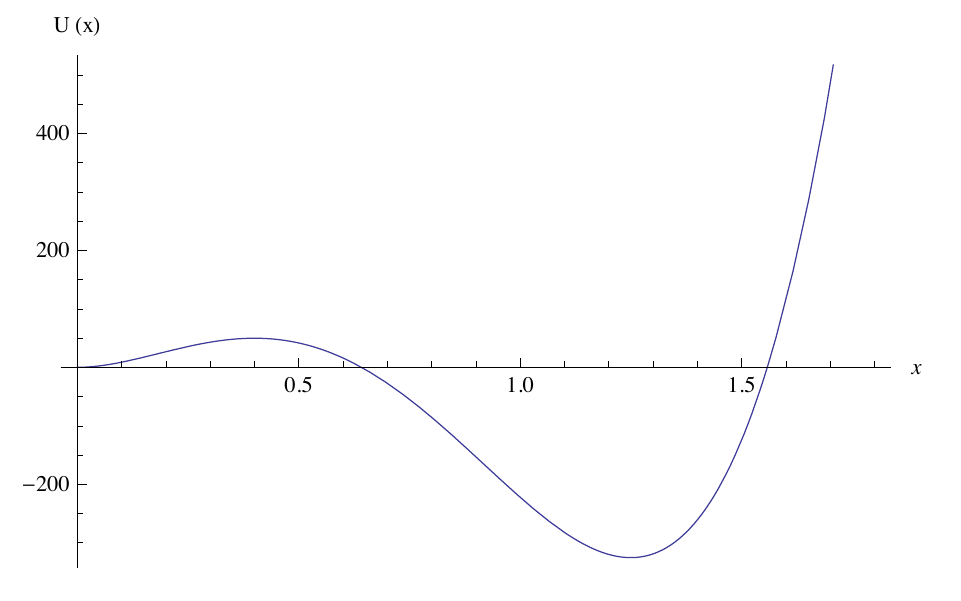}
\caption{The WDW potential $U(x)$ for the parameter values $\lambda=0.03$ and $\gamma = 1.1$.}
\end{center}
\end{figure}
    
For $\gamma =1$, the classically allowed range reduces to a single point, and the WKB action is given by the simple formula
\beq
S_{\gamma = 1} = \frac{1}{6\lambda}.
\label{S1}
\eeq
The classical solution in this case is a static universe with $x=1$,
and Eq.~(\ref{prob}) can be interpreted as being proportional to the
probability of quantum collapse per unit time.  

\section{Tunneling from nothing}

We note that the tunneling between $x=x_-$ and $x=0$ can also go in the
opposite direction, in which case Eq.~(\ref{prob}) with $S$ from (\ref{Sgeneral}) or (\ref{S1}) can 
be interpreted as describing spontaneous creation of an oscillating or static universe
from nothing.  The corresponding instanton can be found by
solving the Euclideanized Friedmann equation,  
\beq
{\dot x}^2 = \omega^2 (x_+ - x)(x_- - x),
\eeq
where the dot stands for differentiation with respect to the Euclidean
time $\tau$.  The solution can be expressed as
\beq
\omega \tau = \int_0^x\frac{dx'}{\sqrt{(x_+-x')(x_--x')}} = -2\ln\left(\frac{\sqrt{x_+ - x}+\sqrt{x_- - x}}{\sqrt{x_+}+\sqrt{x_-}}\right).
\eeq
Solving this for $x$ as a function of $\tau$ we find
\beq
x(\tau)=\gamma-\frac{1}{2}(\gamma-1)e^{\omega\tau}-\frac{1}{2}(\gamma+1)e^{-\omega\tau}.
\label{xtau}
\eeq
Introducing
\beq
\tau_0=\omega^{-1}\ln\left(\frac{\gamma+1}{\gamma-1}\right),
\eeq
Eq.~(\ref{xtau}) can be rewritten as
\beq
x(\tau)=\gamma-\sqrt{\gamma^2-1}\cosh[\omega(\tau-\tau_0/2)].
\label{xtau0}
\eeq
Note that this is related to the Lorentzian solution (\ref{a}) by a
simple analytic continuation, as one might expect.
The instanton solution (\ref{xtau0}) starts with $x=0$ at $\tau=0$, grows until it reaches a maximum value $x(\tau_0/2)=x_-$, and then returns to $x=0$ at $\tau=\tau_0$.  It is symmetric with respect to the point $\tau=\tau_0/2$.

The geometry of the instanton,
\beq
ds^2 = d\tau^2 + a^2(\tau)d\Omega_3^2,
\label{Euclid}
\eeq
is similar to a 4-dimensional ellipsoid.  We note that 
\beq
{\dot a}(0)=-{\dot a}(\tau_0) = 1,
\eeq
which indicates the absence of conical singularities.  In other words, the ``poles'' at 
$\tau=0,\tau_0$ are rounded off.

For $\gamma=1$ the instanton solution (\ref{xtau}) simplifies to
\beq
x(\tau) = 1-e^{-\omega\tau}.
\eeq
It interpolates between $x=0$ at $\tau=0$ and $x=1$ at $\tau\to\infty$.  
The geometry of this instanton is that of a cigar.  It is rounded off
at $a=0$ and asymptotically 
approaches a static sphere at large $\tau$.  The instanton action in
this case is given by
\beq
|S_{inst}|=\frac{3\pi}{4G}\int_0^\infty d\tau a\left[{\dot a}^2+(\omega a
  -1)^2\right] = \frac{1}{6\lambda}.
\eeq
Of course it is the same as in Eq.~(\ref{S1}).  Note that the action
is finite, even though the instanton has an infinite 4-volume.  We
note also that the boundary term, which is proportional to the normal
derivative of the boundary volume, vanishes for this instanton. 

Even though there are no conical singularities, a closer examination shows that somewhat milder singularities are still present at the poles.\footnote{We are grateful to Jaume Garriga for pointing this out to us.}  The scalar curvature for the metric (\ref{Euclid}) is
\beq
R=6a^{-2}(1-{\dot a}^2-a{\ddot a}).
\eeq
The first two terms in the parentheses cancel out at the poles, but in the last term ${\ddot a}(0) = -\gamma\omega\neq 0$, and thus $R\propto a^{-1}\propto\tau^{-1}$.  This singularity is integrable, so the instanton action is finite.  

It is possible that the curvature singularity can be removed by
modifying Einstein's equations or the equation of state at small
values of $a$.  We could imagine, for
example, that for a gas of domain walls the equation of state
parameter gradually changes from $w=-2/3$ to $w=-1$ as we approach
$a=0$ (so the equation of state becomes that of the symmetric vacuum
in the wall interiors).  This would cure the singularity. 

The situation here is somewhat similar to that with the Hawking-Turok (HT) instanton \cite{Turok}, which was proposed to describe quantum creation of open universes.
Garriga has shown that this singular instanton can be regulated with a suitable matter source \cite{Garriga1} and can also be obtained by dimensional reduction from a regular instanton in a higher-dimensional theory \cite{Garriga}.  It is possible that the Euclidean solutions presented here can similarly be regarded as approximations to instantons of a more fundamental theory.

An important difference between HT and our instantons is that in the HT case the vicinity of the singular point makes a significant contribution to the action.  For our instantons the contributions of singular points are negligible.  This indicates that the instanton action and the tunneling probability are not sensitive to short-distance modifications of the theory.

\section{The wave function}

Having studied the semiclassical tunneling of the universe, we shall now examine solutions of the WDW equation (\ref{WDWx}) for the wave function of the universe $\psi(a)$.
By analogy with a quantum harmonic oscillator, one might expect the wave function to oscillate in the classically allowed range and to decay exponentially in the two classically forbidden ranges on both sides of it.  However, the situation we have here is rather different.  In the case of an oscillator, we solve the Schrodinger equation
\beq
\frac{1}{2}\left(-\frac{d^2}{dx^2} + \omega^2 x^2\right) \psi(x)= E\psi(x)
\eeq
with boundary conditions $\psi(x\to\pm\infty)=0$.
Solutions exist only for certain values of the energy, $E=(n+\frac{1}{2})\omega$; this determines the energy spectrum of the oscillator.

Now, in our case the eigenvalue of the WDW operator is fixed: it is
equal to zero.  If we impose boundary conditions requiring, e.g., that
$\psi(x\to\infty) = \psi(x=0) = 0$, the system would be overdetermined
and no solutions would exist, except for some special values of the
parameters $\lambda$ and $\gamma$.  For generic values of the
parameters, we have the freedom to impose only a single boundary
condition.  A natural choice appears to be 
\beq
\psi(x\to\infty)=0.
\label{bc}
\eeq
This fully specifies the solution.  In the classically forbidden region $0\leq x\leq x_-$, the wave function is a superposition of exponentially growing and exponentially decreasing solutions.  The solution that grows towards $a=0$ will dominate, unless the parameters of the model are fine tuned to suppress its contribution.  Some numerical solutions to the WDW equation (\ref{WDWx}) 
are illustrated in Figs.~2 and 3.

\begin{figure}[t]
\begin{center}
\includegraphics[width=12cm]{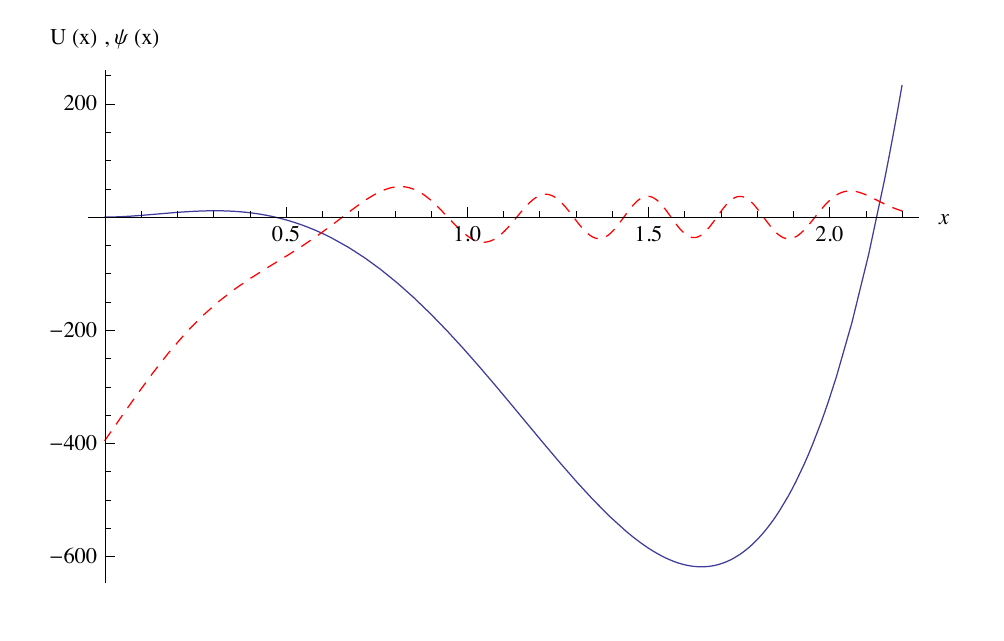}
\caption{Solution of the WDW equation for the parameter values $\lambda = .05$ and $\gamma = 1.3$ (red dashed line).  The WDW potential is also shown (blue line).}
\end{center}
\end{figure}

\begin{figure}[t]
\begin{center}
\includegraphics[width=12cm]{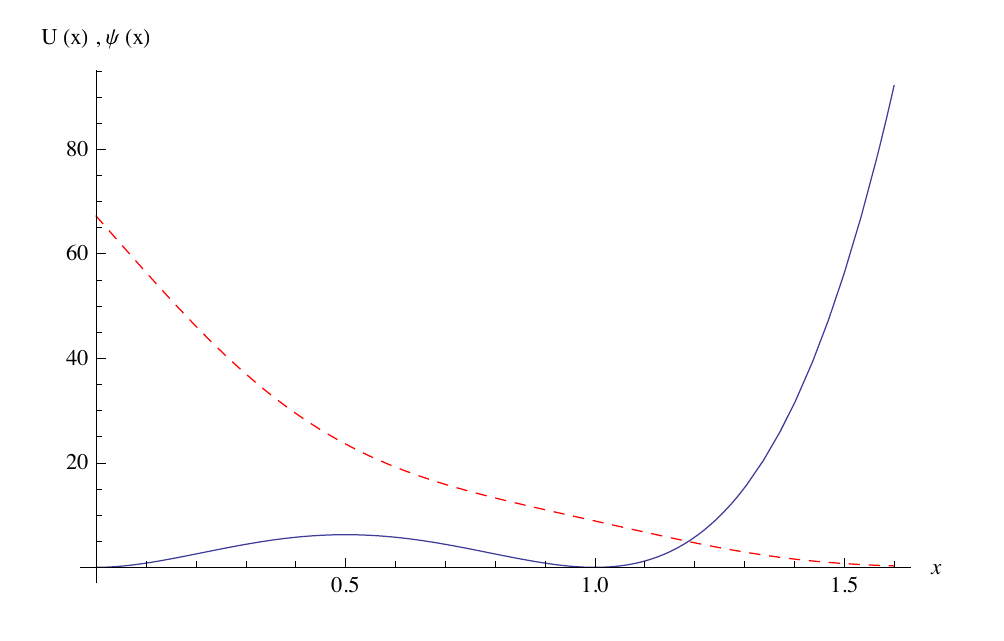}
\caption{Solution of the WDW equation with $\lambda=0.1$ and $\gamma=1$ (red dashed line).  The WDW potential is also shown (blue line).  The classical solution in this case is a static universe at $x=1$.}
\end{center}
\end{figure}

The interpretation of these solutions is not completely clear, since we do not have a well established procedure for extracting probabilities from the wave function of the universe (see, e.g., Ref.~\cite{AV94} and references therein).  But 
a nonzero value of $\psi(0)$ signals a non-vanishing probability of collapse and appears to be  inconsistent with the picture of an eternal oscillating or static universe.  

Here, we assume that hitting the singularity at $x=0$ is fatal for the universe.  It is conceivable that wave functions similar to those in Fig.~3 could describe an eternal universe tunneling back and forth between a finite radius $a=\omega^{-1}$ and a Planck-size nugget.  However, analysis of this possibility would require a full theory of quantum gravity and is beyond our present level of understanding.  Our simple minisuperspace model certainly becomes inadequate at $a\sim l_{Planck} $.

\section{Discussion}
 
Our analysis in this paper indicates that oscillating and static models of the universe, even though they may be perturbatively stable, are generically unstable with respect to quantum collapse. 
Here we focused on the simple harmonic universe with matter content described by Eq.~(\ref{rhosimple}), but we expect our conclusions to apply to a wider class of models.  In particular, one could investigate the quantum stability of braneworld, loop quantum cosmology, and other modified gravity inspired models.\footnote{Some relevant discussion of quantum cosmology in Horava-Lifshitz gravity models can be found in Ref.~\cite{Bertolami}.} 

Is it possible to save the simple harmonic universe from quantum collapse?  One possibility is to impose the boundary condition 
\beq
\psi(0)=0.
\label{psi=0}
\eeq
(This boundary condition was introduced in \cite{DeWitt}; for a recent discussion see \cite{Kiefer2}.)
Together with the boundary condition at infinity (\ref{bc}), this will enforce a relation between the parameters of the model $\gamma$ and $\lambda$.  As Figs.~2 and 3 illustrate, the value of $\psi(0)$ can be either positive or negative.  This is determined by whether $\psi(x)$ is growing or decreasing near $x=x_-$, which is in turn determined by the number of oscillations $N$ of $\psi$ that fit into the classically allowed range $x_-<x<x_+$.  Suppose for definiteness that we decrease $\lambda$ while keeping $\gamma$ fixed.  This makes the potential well deeper, so $N$  monotonically increases and $\psi(0)$ oscillates between positive and negative values, making one oscillation as $N$ changes by $\Delta N\sim 1$.  By continuity, $\psi(0)$ should go through zero twice per such oscillation.  Values of $\lambda\ll 1$ correspond to the semiclassical regime, where $N\gg 1$ and the boundary condition (\ref{psi=0}) can be satisfied by a relatively small change in $\lambda$.

Thus, for each value of $\gamma>1$ we expect an infinite set of values of $\lambda$ for which the condition (\ref{psi=0}) can be enforced.  Fig.~4 shows the wave function for a universe with the parameters fine-tuned in this way.  This approach appears to avoid the collapse, but the following argument indicates that it may not be possible to extend it beyond minisuperspace.

\begin{figure}[t]
\begin{center}
\includegraphics[width=12cm]{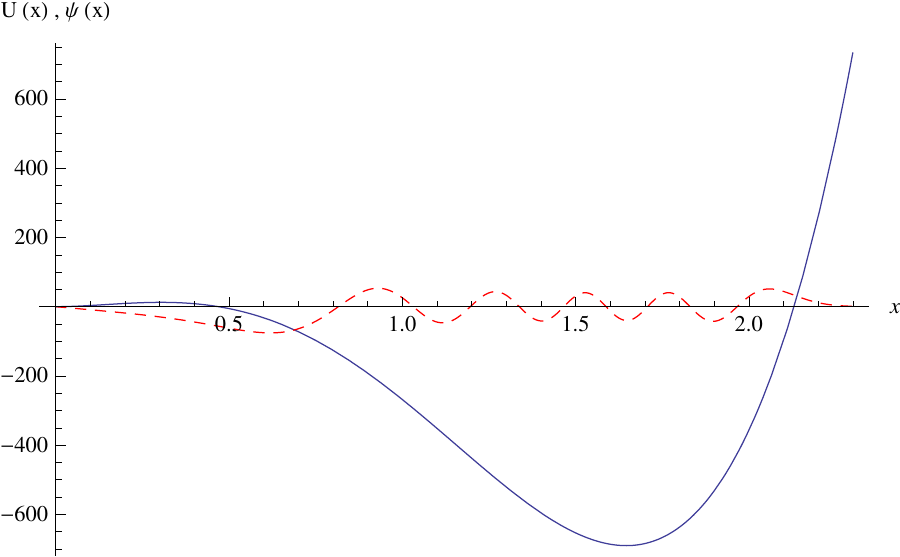}
\caption{Solution of the WDW equation with the parameter values $\lambda=.0473$ and $\gamma=1.3$, fine tuned so that $\psi(0)=0$ (red dashed line).  The WDW potential is also shown (blue line).}
\end{center}
\end{figure}

The WDW equation (\ref{WDWeq}) can be interpreted as stating that the energy of a closed universe is equal to zero.  Quantum states with different occupation numbers of matter particles have different energy of matter, but this energy is exactly compensated by the negative energy of gravity, so the total energy is zero.  Then one expects that transitions between different states should be possible, as long as they have the same conserved quantum numbers.  For example, there seems to be nothing to prevent spontaneous nucleation of particle-antiparticle pairs.  This seems to suggest that the universe will evolve to a state with large occupation numbers and high entropy.  In terms of the wave function, we expect $\psi$ to be a superposition of states with different occupation numbers. 
The value of $\psi(0)$ cannot be fine-tuned for all of them.
Hence, we expect that quantum collapse cannot be prevented by fine-tuning in more realistic models including a quantum matter field.  It would be interesting to study this issue quantitatively. 

A somewhat puzzling aspect of the WDW equation is that the wave function of the universe is independent of time.  Following DeWitt \cite{DeWitt}, we can interpret this as indicating that time should be identified with some semiclassical variable characterizing the universe.  In other words, clocks, being a part of the universe, should also be described by the wave function of the universe.  Our minisuperspace model has a single dynamical variable $a$.  If we use $a$ to measure time, the model has no other variables whose evolution we can describe as a function of time.  Moreover, the radius of the universe $a$ is a rather poor clock in static or oscillating models.  The role of a clock in such a universe can be played by the homogeneous mode of a massless, minimally coupled scalar field, $\phi(t)$.  The field equation for $\phi$ in the metric (\ref{metric}) yields
\beq
a^3{\dot\phi}={\rm const}.
\eeq
This shows that classically $\phi$ changes monotonically with time, which makes it a good clock.
More generally, a clock can be defined whenever the model has some semiclassical variables described by a WKB factor in the wave function, as discussed, e.g., in \cite{DeWitt,Lapchinsky,Banks,Halliwell,AV89}.   With such variables included, it may be possible to quantitatively define such concepts as the tunneling probability per oscillation period or per unit time.
We hope to return to some of these issues in subsequent work.
 


\subsection*{Acknowledgements}

We are grateful to Jose Blanco-Pillado, Jaume Garriga, and Ben Shlaer for very useful discussions and to David Coule and Eduardo Guendelman for useful comments on the manuscript.  This work was supported in part by the National Science Foundation under grant PHY-0855447.


\begin{thebibliography}{99}

\bibitem{Borde}
A. Borde, A.H. Guth and A. Vilenkin, Phys. Rev. Lett. {\bf 90}, 151301 (2003). 

\bibitem{AV82}
A. Vilenkin, Phys. Lett. {\bf 117B}, 25 (1982).

\bibitem{Mulryne} 
D.J. Mulryne, R. Tavakol, J.E. Lidsey and G.F.R. Ellis, Phys. Rev. {\bf D71}, 123512 (2005).

\bibitem{Sergio}
S. del Campo, E. Guendelman, A.B. Kaganovich, R. Herrera and P. Labrana, arXiv:1105.0651 [hep-th].

\bibitem{Yu}
P. Wu and H. Yu, Phys. Rev. {\bf D81}, 103522 (2010).

\bibitem{Dabrowski}
M. P. Dabrowski, Ann. Phys. {\bf 248}, 199 (1996).

\bibitem{Barrow}
J.D. Barrow, G.F.R. Ellis, R. Maartens, and C.G. Tsagas, Class. Quant. Grav. {\bf 20}, L155 (2003).

\bibitem{Graham}
P.W. Graham, B. Horn, S. Kachru, S. Rajendran, and G. Torroba, arXiv:1109.0282 [hep-th]. 

\bibitem{Coule}
D. Coule, Class. Quant. Grav. {\bf 22}, R125 (2005).

\bibitem{Grisha}
C. Barcelo and G.E. Volovik, JETP Lett. {\bf 80}, 239 (2004).

\bibitem{Spergel}
M. Bucher and D.N. Spergel, Phys. Rev. {\bf D60}, 043505 (1999).

\bibitem{DeWitt}
B.S. DeWitt, Phys. Rev. {\bf 160}, 1113 (1967).

\bibitem{AV94}
A. Vilenkin, Phys. Rev. {\bf D50}, 2581 (1994).

\bibitem{Kiefer}
C. Kiefer and B. Sandhofer, arXiv:0804.0672 [gr-qc].

\bibitem{Halliwell}
J.J. Halliwell, in {\it Proceedings of the 1990 Jerusalem Winter School on Quantum Cosmology and Baby Universes}, ed. by S. Coleman, J.B. Hartle, T. Piran and S. Weinberg (World Scientific, Singapore, 1991).
 
\bibitem{DabrowskiLarsen}
M. P. Dabrowski and A. L. Larsen, Phys. Rev. { \bf D52}, 3424 (1995).
 
\bibitem{Turok}
S.W. Hawking and N.G. Turok,  Phys. Lett. {\bf B425}, 25 (1998).


\bibitem{Garriga1}
J. Garriga, Phys. Rev. {\bf D61}, 047301 (2000).
 
\bibitem{Garriga}
J. Garriga, arXiv:9804106 [hep-th]. 

\bibitem{Bertolami}
O. Bertolami and C.A.D. Zarro, arXiv:1106.0126 [hep-th].

\bibitem{Kiefer2}
C. Kiefer, J. Phys. Conf. Ser. {\bf 222}, 012049 (2010).

\bibitem{Lapchinsky}
V. Lapchinsky and V.A. Rubakov, Acta Phys. Pol. {\bf B10}, 1041 (1979).

\bibitem{Banks}
T. Banks, Nucl. Phys. {\bf B249}, 332 (1985).

\bibitem{AV89}
A. Vilenkin, Phys. Rev. {\bf D39}, 1116 (1989).

\end{thebibliography}
\end{document}